\documentclass[aps,pre,twocolumn,showpacs,superscriptaddress,preprintnumbers,amsmath,amssymb,floatfix]{revtex4}
\usepackage{graphicx}
\usepackage{dcolumn}
\usepackage{bm}
\usepackage[T1]{fontenc}
\usepackage[utf8x]{inputenc}
\usepackage{MnSymbol}

\def\<{\langle}
\def\>{\rangle}

\def\Zeta{Z}

\begin{document}

\title{Complexity in surfaces of densest packings for families of polyhedra}

\author{Elizabeth R. Chen}
\thanks{Contributed equally}
\email{bethchen@umich.edu}
\affiliation{School of Engineering and Applied Sciences, Harvard University, Cambridge, MA 02138, USA}
\affiliation{Department of Chemical Engineering, University of Michigan, Ann Arbor, MI 48109, USA}
\author{Daphne Klotsa}
\thanks{Contributed equally}
\email{dklotsa@umich.edu}
\affiliation{Department of Chemical Engineering, University of Michigan, Ann Arbor, MI 48109, USA}
\author{Michael Engel} 
\affiliation{Department of Chemical Engineering, University of Michigan, Ann Arbor, MI 48109, USA}
\author{Pablo F. Damasceno}
\affiliation{Applied Physics Program, University of Michigan, Ann Arbor, MI 48109, USA}
\author{Sharon C. Glotzer}
\email{sglotzer@umich.edu}
\affiliation{Department of Chemical Engineering, University of Michigan, Ann Arbor, MI 48109, USA}
\affiliation{Applied Physics Program, University of Michigan, Ann Arbor, MI 48109, USA}
\affiliation{Department of Materials Science and Engineering, University of Michigan, Ann Arbor, MI, 48109, USA}

\begin{abstract}
Packings of hard polyhedra have been studied for centuries due to their mathematical aesthetic and more recently for their
applications in fields such as nanoscience, granular and colloidal matter, and biology. In all these
fields, particle shape is important for structure and properties, especially upon crowding.
Here, we explore packing as a function of shape. By combining simulations and analytic calculations, 
we study three 2-parameter families of hard polyhedra and report an extensive and systematic analysis of 
the densest packings of more than 55,000 convex shapes. The three families have the symmetries of triangle 
groups (icosahedral, octahedral, tetrahedral) 
and interpolate between various symmetric solids (Platonic, Archimedean, Catalan). 
We find that optimal (maximum) packing density surfaces that reveal unexpected richness and complexity, 
containing as many as 130 different structures within a single family.
Our results demonstrate
the utility of thinking of shape not as a static property of an object in the context of packings, 
but rather as but one point in a higher dimensional shape space whose neighbors in 
that space may have identical or markedly different packings. 
Finally, we present and interpret our packing results in a consistent and generally applicable way 
by proposing a method to distinguish regions of packings and classify types of transitions between them.
\end{abstract}

\pacs{61.50.Ah}

\maketitle


\section{Introduction}

The optimization problem of how to pack objects in space as densely as possible has a long and colorful history~\cite{hales06, bezdek10,aste}. Packing problems are both easy 
to grasp and notoriously hard to solve mathematically, qualities that have made them interesting recreational math puzzles~\cite{demain07}.
Recent work on nanoparticle and colloidal self-assembly~\cite{esco11,arkus11,geissler12}, micrometer molecule analogs~\cite{song2013}, reconfigurability~\cite{yoo2010,maye2010,nguyen2010,nguyen2011,zhang2011,gang2011,lee2012,kohl2013,guo2013},  
and jammed granular matter~\cite{liu10,torq10}, as well as biological cell aggregation~\cite{hayashi2004,astrom2006} and crowding~\cite{Ellis2001,Balbo2013},
has motivated further the study of packing.
Packing in containers has a broad range of applications in operations research, 
such as optimal storage, packaging and transportation~\cite{operations1,operations2}.

Despite significant progress in the study of packing, knowledge remains patchy and focuses on a few selected shapes with high symmetry.
The densest packing of the sphere, known as the Kepler conjecture and formulated over 400 years ago~\cite{kepler1611}, 
was proven by Hales in 2005~\cite{hales05,hales06,lagarias11}.
Besides the sphere and objects that tile space (\emph{i.e.}~fill space completely without gaps or overlaps), no mathematical 
proofs have been found and results are obtained numerically.
Motivated by the great diversity of nanoparticle shapes that can now be synthesized~\cite{geissler12,huang12,huang13}, 
many groups have studied the densest packing of highly symmetric polyhedra 
(Platonic solids, Archimedean solids, and others)~\cite{betke00,chen08,torquato09,chen10,kallus11,dijkstra11,acs,amir13}.
Yet, finding the densest (optimal) packing is challenging even for seemingly simple shapes such as 
the tetrahedron~\cite{chen08,torquato09,kallus10,chen10,kallus11,lagarias12}.
Recent experiments have gone even further by synthesizing nanoparticles --- specifically nanocubes (superballs)~\cite{zhang2011}, 
whose shape can be tuned from a cube to an octahedron via a sphere --- generating homologous families of shapes 
whose packings may vary with shape. To date, most theoretical and/or computational studies have reported the 
densest packings for shape deformations of 1-parameter families (that is, one `axis' in `shape space'). 
Examples include ellipsoids~\cite{donev04}, superballs~\cite{jiao09,zhang2011},
puffy tetrahedra~\cite{kallus-puffy11}, concave $n$-pods~\cite{dijkstra10} and bowls~\cite{dijkstra11}, 
convex shapes characterized by aspect ratios~\cite{ras11}, as well as
truncated polyhedra, such as the tetrahedron-octahedron family~\cite{acs}, the octahedron-cube family~\cite{dijkstra13} 
and tetrahedral dimers~\cite{amir13}. Many of these recent studies report a diversity of densest packings as a function of shape, resulting in a topographically 
complex line through what is actually a high dimensional shape space.
The behavior of higher dimensional maximum density surfaces in shape space obtained by varying two or more 
shape parameters simultaneously, as we do in this paper, affords a more in-depth look at 
the role of shape in packing. In particular, such a study allows for the identification of topographical features (valleys, ridges and tangents) that we define. These definitions facilitate the comparison among different packing studies. 

In this paper, we investigate the packing problem for three 2-parameter families of symmetric convex polyhedra.
Our families interpolate between edge-transitive polyhedra via continuous vertex and/or edge truncations. The interpolation 
goes through various solids (Platonic, Archimedean, Catalan), thereby including some of the above-referenced 1-parameter studies as 
linear paths (subfamilies) on our 2-parameter surfaces.
The densest packings form surfaces in shape space that reveal great diversity in richness and complexity.
Our results demonstrate that some not-previously-studied paths through shape space give a plethora of consecutive distinct packing
structures through a series of transitions (as the shape deforms) whereas other paths give the same or similar packings with no or few transitions. 
Given the richness of the surfaces of densest packings, we aim to standardize the way packing results are presented and interpreted in the community 
by doing the following. Based on the theories of Minkwoski~\cite{mink53,Minkowski1904} and the Kuperbergs~\cite{kuper90}, 
we define regions of topologically equivalent packings from their intersection equations 
(contacts with nearest neighbors) as opposed to Bravais lattice type or symmetry group.
We thus define and classify three types of boundaries between adjacent regions (valley, ridge, tangent) and their combinations.
The classification is general for any convex shape.
We analyze previous works and argue that packing problems can be treated consistently using our framework.

The paper is organized as follows. Section II introduces some of the theoretical concepts and mathematical tools that have been formulated over the years
for packing problems. We construct three 2-parameter families of symmetric polyhedra in section III.
Section IV describes our analytical, numerical, and computational methods.
We show results for the surfaces of maximum packing density in section V and close with  
a comparison with other studies in section VI, and a summary of our main points and conclusion in section VII.


\section{Theoretical background}

The packing problem is an optimization problem that searches for the densest possible packing arrangement of objects ${\it\Xi}$ in a container
or in infinite Euclidean space. It is in general an intractable problem that does not allow a rigorous analytic treatment or numerical search.
When packing in containers, the problem depends on a finite number of 
object positions and orientations and can often be solved via a brute force search~\cite{szabo07}. 
Packing in infinite space requires the optimization of an infinite number of variables.
Here, we focus on packing of identical convex objects (convex particles) in infinite space.
Because all known densest packings of convex objects are periodic, we restrict our search to those.
The packing density of a periodic packing is $\phi=nU/V$,
where $U$ is the volume of the object, $V$ is the volume of the unit cell of the lattice, and $n$ is the number of objects in the unit cell. 

\subsection{Sum and difference bodies}

Given two particles ${\it\Xi}$ and ${\it\Zeta}$, the sum body and the difference body are the sets of sums and differences of all points in the particles:

\medskip\quad ${\it\Xi}+{\it\Zeta} = \{\xi+\zeta : \xi\in{\it\Xi}, \zeta\in{\it\Zeta}\}$,
\par\quad ${\it\Xi}-{\it\Zeta} = \{\xi-\zeta : \xi\in{\it\Xi}, \zeta\in{\it\Zeta}\}$.

\medskip \noindent
For any particle ${\it\Xi}$, we define positive orientation as $+{\it\Xi}={\it\Xi}$ and negative orientation $-{\it\Xi}$ by inversion at the origin.
Two particles ${\it\Xi}$ and ${\it\Zeta}$ are parallel, if there exists a vector $\varsigma$ such that ${\it\Zeta} = \varsigma+{\it\Xi}$.
They are antiparallel, if there exists a vector $\varsigma$ such that ${\it\Zeta} = \varsigma-{\it\Xi}$.
Two particles ${\it\Xi}$ and ${\it\Zeta}$ are in contact, if their intersection 
is equal to the intersection of their boundaries, ${\it\Xi} \cap {\it\Zeta} = \partial{\it\Xi} \cap \partial{\it\Zeta}$.
From basic set theory we know that:

\begin{itemize} 

\item If a convex particle ${\it\Xi}$ (centered at $0$) and a parallel neighbor $\varsigma+{\it\Xi}$ (centered at $\varsigma$) touch, 
then $\varsigma$ lives on the surface of the difference body ${\it\Xi}-{\it\Xi}$.
\item If a convex particle ${\it\Xi}$ (centered at $0$) and an antiparallel neighbor $\varsigma-{\it\Xi}$ (centered at $\varsigma$) 
touch, then $\varsigma$ lives on the surface of the sum body ${\it\Xi}+{\it\Xi}$. 
\item The sum body of particle ${\it\Xi}$ is always convex, even if ${\it\Xi}$ is not.
\item For a convex particle ${\it\Xi}$, the sum body has the same shape but twice the size: ${\it\Xi}+{\it\Xi} = 2{\it\Xi}$.
\item The difference body of particle ${\it\Xi}$ is always centrally symmetric, even if ${\it\Xi}$ is not.
\item For a centrally symmetric particle ${\it\Xi}$, the difference body equals the sum body.
\end{itemize}

\noindent See Fig.~A1 for examples of the sum and difference bodies of non-centrally symmetric polyhedra.

\subsection{Minkowski lattices}

A lattice packing is a packing with one particle in the unit cell ($n=1$).
It has been observed that the densest known packing for many convex particles with
central symmetry is a lattice packing~\cite{torquato09,acs,dijkstra11,amir13}, but this is not 
generally true~\cite{donev04}.
Densest packings of non-centrally symmetric shapes frequently require two or more particles in the unit cell.

The theory of lattice packings was originally developed by Minkowski~\cite{mink53,Minkowski1904}, who
described packings considering the contacts of a particle with its neighbors. 
Chen~\cite{chen-thesis} used this method to study densest packings of various shapes.
In a lattice packing all particles are parallel and have identical neighborhoods related by translations, which are linear
combinations with integer coefficients of the lattice vectors $\{\chi,\psi,\omega\}$.
Minkowski proved that the densest lattice packing (Minkowski lattice) of a particle ${\it\Xi}$ is always identical to the
densest lattice packing of its difference body ${\it\Xi}-{\it\Xi}$. He also proved that for all centrally symmetric convex shapes,
a Minkowski lattice can always be chosen such that each particle ${\it\Xi}$ is in contact with either 12 or 14 neighbors and
the lattice satisfies one of three possible types with either 6 or 7 pairs of neighbor contacts:

\medskip\quad $G^{6-} = \pm\{\chi,\psi,\omega,\psi-\omega,\omega-\chi,\chi-\psi\}$,
\par\quad $G^{6+} = \pm\{\chi,\psi,\omega,\psi+\omega,\omega+\chi,\chi+\psi\}$,
\par\quad $G^{7+} = \pm\{\chi,\psi,\omega,\psi+\omega,\omega+\chi,\chi+\psi,\chi+\psi+\omega\}$.

\medskip 
\noindent The \emph{Minkowski lattice type} $G_{}^{6-}$, $G_{}^{6+}$, or $G_{}^{7+}$ refers to the number and positions of the reference particle contacts with
its neighbors. The Bravais lattice type usually employed in crystallography is not equivalent because it refers to particle centers. 
We can draw a connection if we consider the space-filling packing of Voronoi polyhedra for the face-centered cubic (fcc), 
and the body-centered cubic (bcc) packings, which are the rhombic dodecahedron, and the truncated octahedron respectively.
We can then think of the fcc packing as an example of $G_{}^{6-}$ (with 12 contacts), and the bcc packing
as an example of $G_{}^{7+}$ (with 14 contacts).

\subsection{Kuperberg pairs}

A double lattice packing of a convex particle ${\it\Xi}$ has two antiparallel particles in the unit cell ($n=2$),
$+{\it\Xi}$ and $\delta-{\it\Xi}$, which form a \emph{Kuperberg pair}~\cite{kuper90}.
The Kuperbergs extended Minkowski's theory to double lattice packings. 
They showed that the pair often packs densely for particles without central symmetry and is a candidate for the solution of the 
general packing problem in situations when lattice packings are not a good solution.
Chen \emph{et al.}~\cite{chen10,chen-thesis} and Haji-Akbari \emph{et al.}~\cite{amir13} used this method as well in their studies
of densest packings. 

Exceptions where densest packings are realized in neither the Minkowski lattice nor a Kuperberg pair are known.
For example, the densest packing of ellipsoids has $n=2$, but is not a double lattice packing because the two ellipsoids in the unit cell 
are not antiparallel~\cite{donev04}.
The densest packing of tetrahedra requires $n=4$~\cite{chen10,acs,kallus11,amir13}. 
In two dimensions, space-filling packings (tilings) of pentagons are known with $n=2,3,4,6,8$~\cite{grunbaum}.


\section{Construction of 2-parameter families of polyhedra}
\label{sec:meet-families}

Because the maximum packing density is a function of the geometric shape of the particles, it is useful to describe continuous deformations of particle shape ${\it\Xi}$.
An $N$-parameter family of three-dimensional shapes is a function $F: {\bf R}_{}^N\rightarrow{\bf R}_{}^3$.
The parameters $X = \<X_1^{},\cdots,X_N^{}\>\in{\bf R}^N$ then represent specific operations on the shape ${\it\Xi} = F(X)$.

\subsection{Spheric triangle groups}

We introduce families of polyhedra that interpolate between various symmetric solids (Platonic, Archimedean, Catalan) via truncation.
The amount of truncation is varied in each family by shifting the truncation planes radially in a manner respecting a centrosymmetric point symmetry group.
Such symmetry groups are known as finite spheric triangle group $\Delta_{p,q,r}^{}$.
A spheric triangle group is generated by three reflections across the sides of a spheric triangle with angles $\{{\pi\over p},{\pi\over q},{\pi\over r}\}$.
The finite irreducible spheric triangle groups are $\Delta_{3,2,3}$ (tetrahedral, Sch\"onflies notation T$_\text{d}$), $\Delta_{4,2,3}$ (cubic-octahedral, O$_\text{h}$), and
$\Delta_{5,2,3}$ (dodecahedral-icosahedral, I$_\text{h}$)~\footnote{We exclude the irreducible spheric triangle groups $\Delta_{p,2,2}$ (axial-dihedral, D$_{p\text{h}}$)
because its intermediate polyhedra (polygonal prisms, polygonal bipyramids) are very anisotropic.}.

\subsection{Truncation planes}

\begin{figure*}
\centering
\includegraphics[width=6.5in]{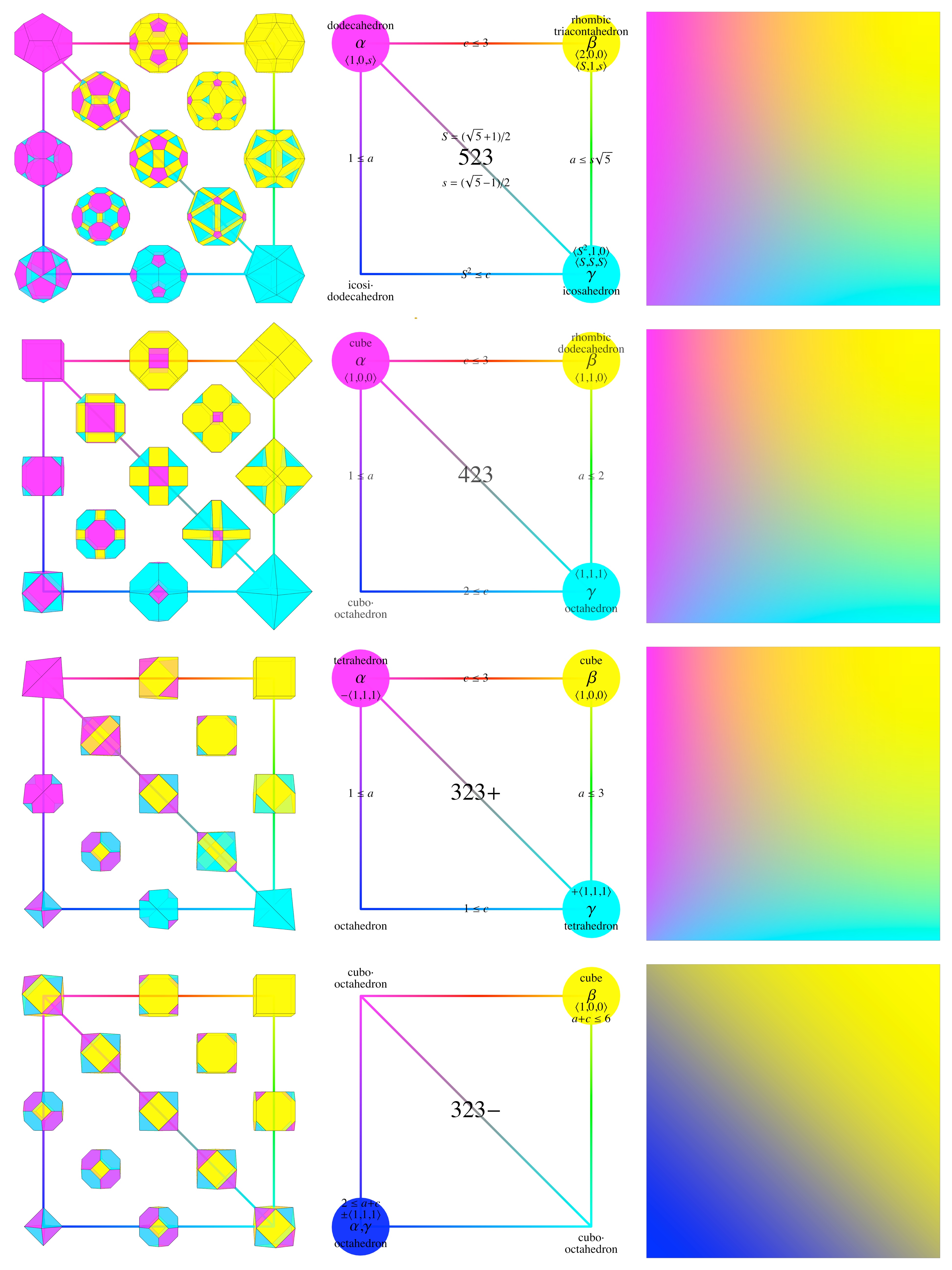}
\caption{Left column: Representative polyhedra for the families $523$, $423$, $323+$, $323-$. The two pairs of diagonally opposite corners correspond to dual polyhedra.
The $323$ family also has reflection symmetry about the diagonal $a = c$. Middle column: The direction of face normal vectors $\{\alpha,\beta,\gamma\}$,
parameter range $\<a,c\>$, and the names of the corner polyhedra. Right column: A color map of the surface area fraction of face types.
In all columns, the color $\{{\rm magenta},{\rm yellow},{\rm cyan}\}$ indicates the type of face $\{\alpha,\beta,\gamma\}$.}
\label{fig:familyalbum}
\end{figure*} 

Three 2-parameter families of polyhedra are constructed by truncating the vertices and edges of the dodecahedron or icosahedron ($523$ family),
the cube or octahedron ($423$ family), and the tetrahedron ($323$ family).
The result are three types of equivalent face normal vectors $\{\alpha,\beta,\gamma\}$ that are axes of $\{p,q,r\}$-fold symmetry 
(rotation by angles $\{{2\pi\over p},{2\pi\over q},{2\pi\over r}\}$).
The triangle group maps any axis to any other axis of the same type.
We define the polyhedron ${\it\Xi}$ as the intersection of half spaces for all normal vectors,
where the parameters $\{a,b,c\}$ specify the amount of truncation or position of the bounding plane:

\medskip\quad ${\it\Xi}$ $\begin{cases}
\xi\cdot\alpha\le a\\
\xi\cdot\beta\le b\\
\xi\cdot\gamma\le c\end{cases}$

\medskip \noindent The normal vectors and parameter ranges for the three families are given by

\medskip\quad  523 $\begin{cases}
\hbox to 60pt{$\alpha = \<1,0,s\>$\hfill} 1\le a\le s\sqrt 5\\
\hbox to 60pt{$\beta = \<2,0,0\>$\hfill} 2 = b\\
\hbox to 60pt{$\gamma = \<S,S,S\>$\hfill} S_{}^2\le c\le 3\end{cases}$
$\begin{cases}
S = {1\over 2}(\sqrt 5+1)\\
s = {1\over 2}(\sqrt 5-1)\end{cases}$

\medskip\quad 423 $\begin{cases}
\hbox to 60pt{$\alpha = \<1,0,0\>$\hfill} 1\le a\le 2\\
\hbox to 60pt{$\beta = \<1,1,0\>$\hfill} 2 = b\\
\hbox to 60pt{$\gamma = \<1,1,1\>$\hfill} 2\le c\le 3\end{cases}$

\medskip\quad 323 $\begin{cases}
\hbox to 60pt{$\alpha =- \<1,1,1\>$\hfill} 1\le a\le 3\\
\hbox to 60pt{$\beta = \<1,0,0\>$\hfill} 1 = b\\
\hbox to 60pt{$\gamma = +\<1,1,1\>$\hfill} 1\le c\le 3 \end{cases}$

\medskip We fix $b$ to be constant because the packing density remains invariant if we rescale all parameters by the same scalar.
Fig.~\ref{fig:familyalbum} shows representative polyhedra on the square $\<a,c\>$ parameter domain.
Note that at the boundaries of the square domain, the polyhedra have degeneracies 
(faces with zero area, edges with zero length, coincident points).
If we include the degeneracies, all family members have the same number of faces, edges, and vertices.
If we exclude the degeneracies, the polyhedra at the corners of the domain are edge-transitive.
Formulae for the polyhedron volume $U$ and face areas $\{\alpha,\beta,\gamma\}$ are given in the appendix.

\subsection{Sum and difference bodies}

Most polyhedra in the $pqr$ family have a point symmetry group that is identical to the triangle group $\Delta_{p,q,r}$.
The only exceptions occur in the $323$ family, where the polyhedra with central symmetry 
are also members of the $423$ family, so they have the higher $\Delta_{4,2,3}^{}$ symmetry.
In fact, $323$ is the only family with non-centrally symmetric polyhedra.
We derive two families: (i)~the family of sum bodies $323+$ which is identical to $323$, and (ii)~the family of difference bodies $323-$
which corresponds to $323$ along the diagonal $a=c$ and does not change in the orthogonal direction $a=-c$.

The sum and difference bodies of polyhedra in the $523$ and $423$ families with central symmetry are identical:

\medskip\quad ${\it\Zeta} = 2{\it\Xi}$ $\begin{cases}
\zeta\cdot\alpha\le 2a\\
\zeta\cdot\beta\le 2b\\
\zeta\cdot\gamma\le 2c\end{cases}$

\medskip \noindent For the 323 family without central symmetry the sum ($323+$) and difference bodies ($323-$) are different:

\medskip\quad ${\it\Zeta} = {\it\Xi}+{\it\Xi}$ $\begin{cases}
\zeta\cdot\alpha\le a+a\\
\zeta\cdot\beta\le b+b\\
\zeta\cdot\gamma\le c+c\end{cases}$

\medskip\quad ${\it\Zeta} = {\it\Xi}-{\it\Xi}$ $\begin{cases}
\zeta\cdot\alpha\le a+c\\
\zeta\cdot\beta\le b+b\\
\zeta\cdot\gamma\le c+a\end{cases}$


\section{Methods}
For each family, we analyze densest packings in two steps.
First, we generate dense packings using Monte Carlo simulations by compressing a small number of $n$ particles with periodic boundary conditions.
We then use these results as a guide to construct an analytic surface of maximum packing density (as in~\cite{amir13} for one shape parameter). 

For polyhedra with central symmetry (523, 423, and $323-$ families), we investigated only lattice packings, 
because the densest known packings of centrally symmetric shapes are most likely found to be lattice packings~\cite{torquato09,acs,dijkstra11,amir13}.
For polyhedra without central symmetry ($323+$ family), we studied packings with $n=1,2,3,4$ particles in the unit cell.
In the following we use the notation $323\!\cdot\!n$ for a packing of a shape in the $323$ family with $n$ particles in the unit cell.

\subsection{Simulated compression with Monte Carlo}

Our computational techniques closely follow previous works~\cite{amir09,acs,dama12,amir13}.
We study small systems of $n$ identical polyhedra in a box of volume $V$ with periodic boundary conditions.
The polyhedra positions and orientations evolve in time according to a Monte Carlo trial move update scheme,
where polyhedra are chosen randomly and then rotated and translated by a random amount.
In addition, the simulation box is updated in the isobaric-isotension ensemble by randomly perturbing the coordinates of the three box vectors.
Strong elongations of the box vectors are avoided by using a lattice reduction technique.
Trial moves are accepted if the generated configurations are free of overlaps and rejected otherwise.
Overlap checks are performed using the GJK algorithm~\cite{gjk88}.
In contrast to previous works, here, self-overlaps are accounted for with periodic copies due to the small dimensions of the simulation box.
Compared to other compression techniques in the literature~\cite{torquato09,kallus10b,dijkstra11,marcotte13}, our scheme consistently 
finds equivalent or denser packings.

For each of the three families, we choose polyhedra from a fine $101\times 101$ parameter grid in the $\<a,c\>$ parameter domain.
In a certain parts of the domain, where the packing types changes rapidly with parameters $a,c$, we apply a finer grid to achieve a higher resolution.
Although there is some overlap between the shapes of the $423$ and $323$ families, 
the total number of unique shapes simulated for this study was more than 55,000.
The simulation is initialized at low density and then slowly compressed by gradually increasing the pressure using an exponential protocol over $7\times10^5$ steps.
Because our approach resembles the simulated annealing technique replacing temperature with pressure, we call it `simulated compression'.
Each compression run is repeated 10 times and the densest packing recorded for each parameter choice.
The result of the algorithm is a numerical candidate for the densest packing function over the two shape parameters.

\subsection{Analytic optimization}

We use the densest packings from simulated compression as a guide to analytically construct small unit cell packings
that are locally optimal under rotations and translations of the particles.
As in the simulated compression simulations, we consider only lattice packings for the $523$ and $423$ families
and investigate unit cells with up to four particles for the $323$ family.

For a lattice packing ($n=1$), we perform the following:
\begin{enumerate}
\item Analyze the neighbor contacts in the densest packings obtained with simulated compression. This step requires some manual work, but is usually straight-forward.
\item Write the (abstract) intersection equations in terms of the
lattice basis vectors $\{\chi,\psi,\omega\}$ and the polyhedron faces, edges, and vertices which are functions of the shape parameters $\<a,c\>$.
\item Reduce the parameters $\{\chi,\psi,\omega,a,c\}$ to a minimal set of free parameters.
\item If the (abstract) lattice volume $V = \det[\chi,\psi,\omega]$ has free parameters, find the values of the free parameters that minimize $V$ and
therefore maximize the packing density $\phi = U/V$.
\end{enumerate}
In general, for a given lattice, there are multiple ways to choose a set of basis vectors that generates the lattice.
For centrally symmetric shapes, we choose basis vectors to satisfy one of the three
Minkowski types. This simplifies the optimization procedure and introduces consistency.

For the double lattice packings ($n = 2$) in $323\!\cdot\!2$, we repeat the same process as for the lattice packing with the offset vector $\varsigma$ as an additional parameter. 
By coincidence, the $n=4$ case ($323\!\cdot\!4$ family) reduces to double lattice packings.
The only exception is an area near the tetrahedra ($\<a,c\>=\<1,3\>$ and $\<3,1\>$), which is a double lattice packing of dimers only slightly rotated
(face-face, almost edge-edge).
Note that we use a packing of two monomers, whereas the Kuperbergs~\cite{kuper90} use a packing of one dimer (Kuperberg pair)
and thus there are additional degrees of freedom in our packing.
The packing of the $323\!\cdot\!2$ family (and most of the $323\!\cdot\!4$ family) is a combination of the sum body $323+$ and 
difference body $323-$ packings, in that the neighbors in the densest packing are either parallel or antiparallel.

Because the particles in Kuperberg pairs are related to one another by simple inversion, all intersection equations in the $n=1,2,4$ cases
are linear and can be solved analytically. However, for arbitrary shapes (\emph{e.g.}~truncated triangular bipyramids~\cite{amir13}) and $n\ge 3$
this is not generally true. Rotations between neighboring particles can result in a system of quadratic intersection equations.
This is the case also for the $323\!\cdot\!3$ family, for which we have to solve the nonlinear intersection equations numerically.

The intersection equations of all analyzed families are summarized in the Appendix.
Numeric versus analytic highest packing densities are compared in Fig.~A3.

The simulated compression results for the surface of maximum packing density, it is easy to miss small regions.
We try to mitigate this problem by using a fine resolution especially in areas where we expect complexity (many small regions).
We verify that (i)~the packing density data from simulated compression is always a lower bound to the analytic results, and
(ii)~adjacent regions match up correctly.

\subsection{Classification of packings}

We classify densest packings based on the types of contacts between neighboring particles in the unit cell. 
There are four topological types of contact: face-to-face, edge-to-face, vertex-to-face,
and edge-to-edge.  Each contact is mathematically expressed in the form of an intersection equation.
We refer to packings with the same topological types of contacts and intersection equations as \emph{topologically equivalent}.
Note that the contacts map to each other other via isometry between lattice vectors and/or isomorphism between basis vectors.
The equivalence relation partitions the domain of each $N$-parameter family of packings into equivalence classes, 
which we call \emph{packing regions} and denote by the
symbol $\rho_i^{}$. 
Within each region the packing density varies continuously and smoothly with shape parameters $X$.
This is because the intersection equations and therefore the lattice vectors are smooth (algebraic) functions of the shape parameters $X$.  
At the boundaries between adjacent regions the packing density might be nonsmooth (discontinuous derivative).
The packing density $\phi(X)$ is necessarily continuous everywhere on the domain.

\begin{figure*}
\centering
\includegraphics[width=6.75in,height=2.25in]{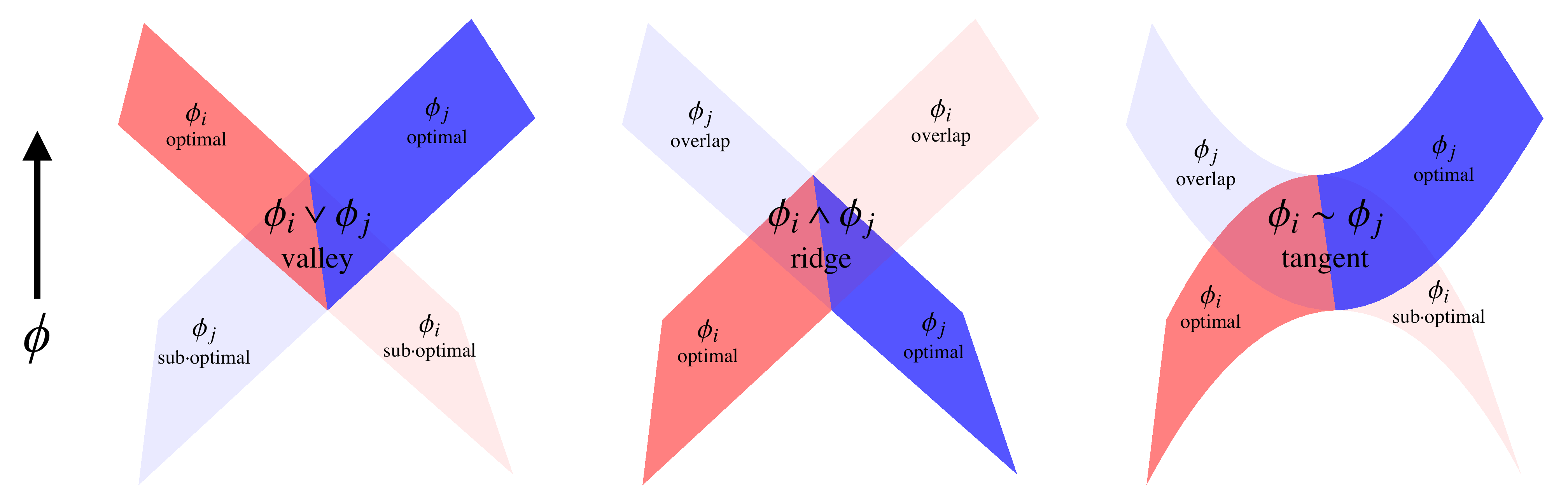}
\caption{Boundaries between adjacent regions of topologically distinct packings are classified into three boundary types:
valley (left), ridge (middle), and tangent (right).}
\label{fig:boundary}
\end{figure*} 

When crossing between adjacent regions, a minimal set of intersection equations changes.
This minimal set depends on the symmetries of adjacent packings and number of particles in the unit cell.
We use the discontinuities of the first derivative of $\phi(X)$ to distinguish three \emph{boundary types} as depicted for a 2-parameter family of packings in Fig.~\ref{fig:boundary}. We define:

\begin{enumerate}
\item A \emph{valley} $[\phi_i^{}\vee\phi_j^{}]$ is a `soft' boundary. 
Extending beyond the valley is allowed but gives sub-optimal packings.
A valley is a generic boundary; intersection equations of adjacent regions 
need not be related, so the lattice vectors can change discontinuously.
We can write the optimal density for a valley as:
\begin{equation} \label{eq:valley}
\phi_{i\vee j}^{} = [\phi_i^{}\vee\phi_j^{}] = \max\{\phi_i^{},\phi_j^{}\} = \begin{cases}
\phi_i^{} & \phi_i^{}\ge\phi_j^{}\\
\phi_j^{} & \phi_i^{}\le\phi_j^{}\end{cases}.
\end{equation}

\item A \emph{ridge} $[\phi_i^{}\wedge\phi_j^{}]$ is a `hard' boundary. 
Extending beyond a ridge introduces overlaps into the packing, which is not allowed.
The lattice volume and lattice vectors vary continuously but not smoothly across a ridge.
Right on the ridge the set of contacts is the union of the sets of contacts on both sides.
Thus the set of intersection equations is the union of the sets of intersection equations on both sides. 
This means there are more contacts on the ridge
than on either side. We can think of a ridge as a lower-dimensional region in its own right.
For a ridge boundary:
\begin{equation} \label{eq:ridge}
\phi_{i\wedge j}^{} = [\phi_i^{}\wedge\phi_j^{}] = \min\{\phi_i^{},\phi_j^{}\} = \begin{cases}
\phi_i^{} & \phi_i^{} \le \phi_j^{}\\
\phi_j^{} & \phi_i^{} \ge \phi_j^{}\end{cases}.
\end{equation}

\item A \emph{tangent} $[\phi_i^{}\sim\phi_j^{}]$ $(\phi_i^{} \le \phi_j^{})$ or $[\phi_i^{}\backsim\phi_j^{}]$ $(\phi_i^{} \ge \phi_j^{})$
is a hybrid between a valley and a ridge.
Extending from the lower density side to the higher density side is allowed, but gives sub-optimal packings.
Extending from the higher density side to the lower density side introduces overlaps.
The lattice volume and lattice vectors are continuous and smooth across a tangent.
The intersection equations of the higher density region are a subset of the intersection equations of the lower density region,
and in this case more constraints results in lower packing density.
For a tangent boundary, we take partial derivatives in any direction transverse to the tangent boundary, 
from the $\phi_i^{}$ side to the $\phi_j^{}$ side:\begin{eqnarray} \label{eq:tang1}
\phi_{i\sim j}^{} &=& [\phi_i^{}\sim\phi_j^{}] = \begin{cases}
\phi_i^{} & \partial\phi_i^{} \ge \partial\phi_j^{}\\
\phi_j^{} & \partial\phi_i^{} \le \partial\phi_j^{}\end{cases},\\
\phi_{i\backsim j}^{} &=& [\phi_i^{}\backsim\phi_j^{}] = \begin{cases}
\phi_i^{} & \partial\phi_i^{} \le \partial\phi_j^{}\\
\phi_j^{} & \partial\phi_i^{} \ge \partial\phi_j^{}\end{cases}.
\end{eqnarray}
\end{enumerate}


\section{Results}

The results from simulated compression and analytic optimization are presented in Figs.~\ref{fig:pack523}-\ref{fig:pack3234}.
In each figure, the top $3\times 3$ grid of images show the surface of maximal packing density $\phi(a,c)$, ${18\over 49}\le\phi\le 1$ in 3D space $\<a,c,\phi\>$.
The normal vector of the surface of maximum packing density maps to the color sphere: 
white along the north pole ($\phi$-direction), bright colors along the equator.
The central cube has perspective from the $+\phi$ direction. The peripheral cubes are rotated by a zenith angle 
from the central cube (${\pi\over 3}$ from the $+\phi$ direction). The eight peripheral cubes have perspective from 
eight equally spaced azimuthal angles (multiples of ${\pi\over 4}$ in the $\<a,c\>$ plane).

The bottom row shows a top view on the $\<a,c\>$ plane of the packing regions (separated by solid lines). 
The colors for each region interpolate between the three colors 
that correspond to the face types in Fig.~\ref{fig:familyalbum} $\{{\rm magenta},{\rm yellow},{\rm cyan}\}$.
On the bottom left, if applicable, the numbers indicate the Minkowski lattice type.
The colors indicate which types of faces are in contact between parallel neighbors only.
On the bottom center, the numbers label the regions in order of area size, $0$ being the smallest.
The colors indicate the proportion of each type of face over the shape's total surface area, visualizing the dominant faces.
Finally, if the polyhedra are not centrally symmetric, there is a bottom right panel.
The colors indicate which types of faces are in contact between antiparallel neighbors only. 

\subsection{523 results}

\begin{figure*}
\centering
\includegraphics[width=6.75in,height=9in]{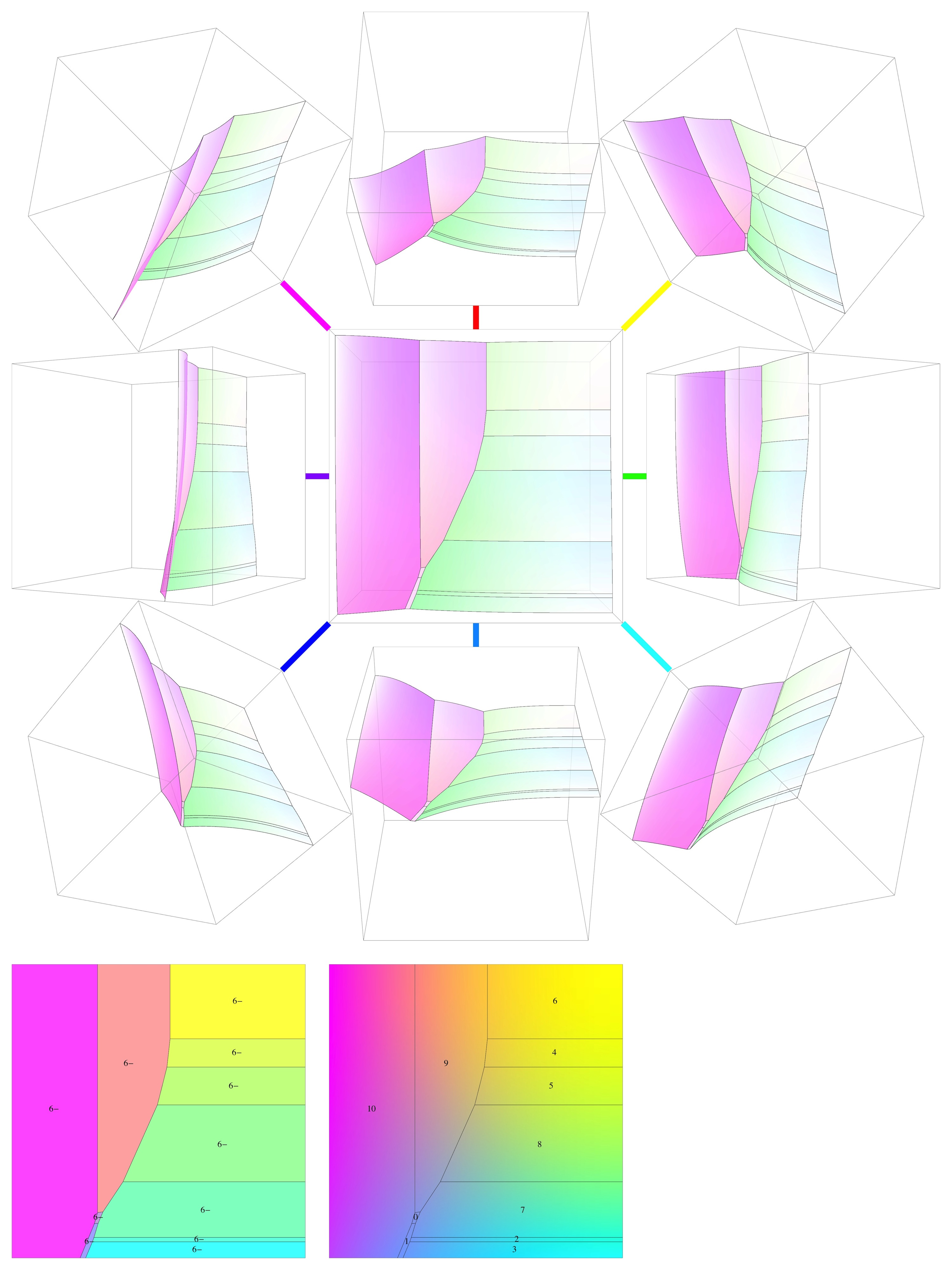}
\caption{$523$ family optimal density surface ($3\times 3$ grid of cubes), parallel contacts (bottom left), face areas and region size (bottom center).}
\label{fig:pack523}
\end{figure*}

Fig.~\ref{fig:pack523} shows the surface of maximum packing density for the 523 family. 
There are 11 regions $\{\rho_i\}$, all with Minkowski lattice type $G_{}^{6-}$, which includes fcc.
All boundaries are valleys, so the overall packing density function is simply the maximum of the functions for each region,
\begin{align*}
\phi = \max\{&\phi_{10}^{6-},\phi_9^{6-},\phi_8^{6-},\phi_7^{6-},\phi_6^{6-},\\
&\phi_5^{6-},\phi_4^{6-},\phi_3^{6-},\phi_2^{6-},\phi_1^{6-},\phi_0^{6-}\},
\end{align*}
where we enumerate the regions and indicate the Minkowski lattice type as superscript.

Consider the colors in the bottom left panel and the numbers in the bottom center.
Along the sequence of regions $\{\rho_6^{}, \rho_4^{}, \rho_5^{}, \rho_8^{}, \rho_7^{}, \rho_2^{}, \rho_3^{}\}$, contacts of yellow $\{\beta\}$ faces 
between neighbors (which are perpendicular to the 2-fold axes) are gradually replaced by contacts of cyan $\{\gamma\}$ faces (3-fold axes).
The same is true for the sequence $\{\rho_6^{}, \rho_9^{}, \rho_{10}^{}\}$, where contacts of yellow faces are gradually replaced by 
contacts of magenta $\{\alpha\}$ faces
(5-fold axes). There is a close similarity between the transitions in colors characterizing the face contact (bottom left panel) 
and the colors characterizing the face area (bottom center panel).
This demonstrates that the face type with the largest area dominates the occurrence of contacts between 
neighbors in the dense packings of the 523 family.

\subsection{423 results}

\begin{figure*}
\centering
\includegraphics[width=6.75in,height=9in]{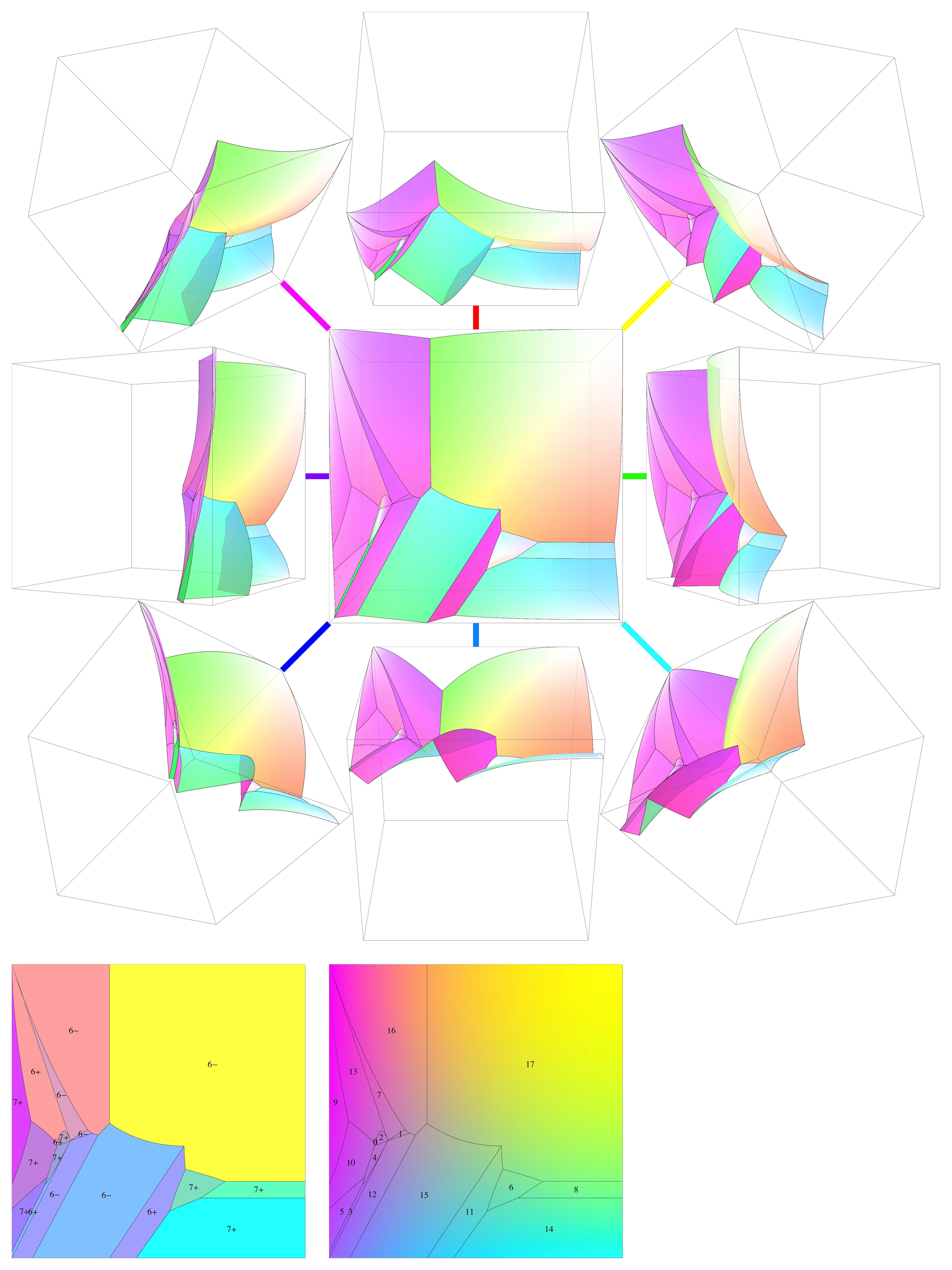}
\caption{$423$ family optimal density surface ($3\times 3$ grid of cubes), parallel contacts (bottom left), face areas and region size (bottom center).}
\label{fig:pack423}
\end{figure*}

Fig.~\ref{fig:pack423} shows the surface of maximum packing density for the 423 family. 
We find 18 regions enumerated in the bottom center panel. 
This is a complex surface, where seven regions meet at the point $\<a,c\> = \<{6\over 5},{12\over 5}\>$.
All boundaries are valleys except for three ridges $[\phi_{15}^{6-}\wedge\phi_{11}^{6+}]$, $[\phi_{13}^{6+}\wedge\phi_7^{6-}]$, $[\phi_{12}^{6-}\wedge\phi_3^{6+}]$,
and one tangent $[\phi_2^{7+}\sim\phi_0^{6+}]$.
We can combine regions so that they all intersect at valleys, thus the overall packing density function is simply 
the maximum of the packing density functions for each region:
\begin{align*}
\phi = \max\{&\phi_{15\wedge 11}^{},\phi_{13\wedge 7}^{},\phi_{12\wedge 3}^{},\phi_{2\sim 0}^{},\\
&\phi_{17}^{6-},\phi_{16}^{6-},\phi_{15}^{7+},\phi_{10}^{7+},\phi_9^{7+},\phi_8^{7+},\phi_6^{7+},\phi_5^{7+},\phi_4^{7+},\phi_1^{6-}\}.
\end{align*}
Unlike the $523$ family, here the colors in the bottom left panel (characterizing the face contact) do not always follow the colors in the bottom center panel
(characterizing the face area). 
Minkowski adjacent colors transition in the same direction as facet area adjacent colors across valleys, and in the opposite direction across ridges.
An example is the sequence $\{\rho_{14}^{}, \rho_{11}^{}, \rho_{15}^{}\}$.
While the boundary between $\rho_{14}^{}$ and $\rho_{11}^{}$ is a valley and the color changes from cyan towards purple in both panels,
$\rho_{15}^{}$ has more cyan than $\rho_{11}^{}$ in the bottom left panel. This means that parts of the magenta faces that were in contact in
$\rho_{11}^{}$ are no longer in contact in $\rho_{15}^{}$, despite the magenta faces being larger in $\rho_{15}^{}$. 

Furthermore, we observe that at a ridge boundary the Minkowski lattice type always changes from $G_{}^{6-}$ to $G_{}^{6+}$.
We can rationalize this behavior in the following way.
By definition a ridge is the boundary where if we extend beyond from either side, the packing has overlaps. This implies that as the shape is deformed,
exactly on the ridge a new contact is introduced. That new contact will become an overlap if we extend beyond the ridge. 
There are only three possible types of lattice packings ($G_{}^{6-}$, $G_{}^{6+}$, $G_{}^{7+}$), 
and two of them ($G_{}^{6-}$ and $G_{}^{6+}$) have the same number of contacts. 
Thus the only way that regions can intersect via a ridge is if the packings have type $G_{}^{6-}$ on one side, type $G_{}^{6+}$ on the other side,
and type $G_{}^{7+}$ at the boundary. 

\subsection{323 family}

Fig.~\ref{fig:pack3231} analyzes the surface of maximum packing density for the $323\!\cdot\!1$ family.
There are eight regions of all three Minkowski lattice types, symmetric about the diagonal $a = c$.
All boundaries are valleys except for two ridges $[\phi_6^{6-}\wedge\phi_1^{6+}]$ and $[\phi_4^{6-}\wedge\phi_0^{6+}]$.
By combining regions separated by ridges we can write down the overall packing density function simply as 
the maximum of the packing density function for each region:
\begin{align*}
\phi &= \max\{\phi_{6\wedge 1}^{},\phi_{4\wedge 0}^{},\phi_7^{7+},\phi_5^{7+},\phi_3^{7+},\phi_2^{7+}\}.
\end{align*}
As required, ridges separate regions of Minkowski lattice types $G_{}^{6-}$ and $G_{}^{6+}$ and have type $G_{}^{7+}$ at the boundary. 
The symmetry about the diagonal is explained by the theory of Minkowski lattices outlined earlier. Since polyhedra in the $323=323+$ family are not centrally symmetric,
their Minkowski lattices are identical to the Minkowski lattices of the family of difference bodies, $323-$. We therefore have to analyze contact and face areas for polyhedra in
the $323-$ family when coloring the bottom left and bottom center panel.

Fig.~\ref{fig:pack3233} analyzes the surface of maximum packing density for the $323\!\cdot\!3$ family.
We have not constructed the analytic surface because there are variable rotations
among the three particles in the unit cell that give quadratic (nonlinear) intersection equations. 
The figure gives an impression of the noise in the raw numerical data and thus demonstrates the significance of the analytic optimization
for obtaining a clean surface of maximum packing density.
The comparison of the results for $323\!\cdot\!1$ and $323\!\cdot\!3$ show that three particles in the unit cell can pack denser than
lattice packings.

Fig.~\ref{fig:pack3232} analyzes the surface of maximum packing density for the $323\!\cdot\!2$ family.
The surface is remarkably complex and has reflection symmetry about the diagonal $a = c$.
Since all packings are double-lattice packings, we can solve for the intersection equations analytically.
There are $8\times 1+62\times 2 = 132$ regions, taking into account reflection symmetry.
Eight regions straddle the diagonal and 62 mirror pairs are reflected across the diagonal.
All boundaries are valleys except 13 ridges and 11 tangents.
We find multiple junctions where several ridges or several tangents meet at a point. 

Fig.~\ref{fig:pack3234} analyzes the surface of maximum packing density for the $323\!\cdot\!4$ family.
There are $8\times 1+61\times 2 = 130$ regions, eight that straddle the diagonal and 61 mirror pairs reflected across the diagonal.
All boundaries are valleys except 13 ridges and 11 tangents, which are the same as for the $323\!\cdot\!2$ above. 

On the symmetry axis $a=c$, which is identical to the subfamily of $323-$ polyhedra with central symmetry, all $323$ families
have the same surface of maximum packing density and the same packing, which is the Minkowski lattice packing.
Near the symmetry axis, in the regions straddling it, all $323$ families have the same density function, but possibly different packings.
The densest packings for $323\!\cdot\!2$ and $323\!\cdot\!4$ are identical except 
for a small area near the tetrahedron corners $\<a,c\>=\<3,1\>$ and its mirror image $\<1,3\>$,
where the regions $\rho_{60}^{}$ and $\rho_{34}^{}$ consist of a Kuperberg pair of dimers. 


\section{Comparison with other packing studies}

Packing studies of 1-parameter families have observed some of the maximum density surface topography presented
here~\cite{donev04,jiao09,torquato09,kallus-puffy11,dijkstra11,acs,amir13,dijkstra13}.
In this section, we apply our classification of regions of packings and boundaries 
to earlier works and compare their analyses of the maximum density graphs to our surfaces of maximum density. 

In Fig.~3 of Ref.~\cite{kallus-puffy11} on puffy tetrahedra, Kallus and Elser describe four regions and three transitions,
two of them as abrupt (between $D_1^{}$ and $S_1^{}$, and $S_1^{}$ and $D_0^{}$) 
and one as continuous (between $S_0^{}$ and $D_1^{}$). 
From their data we would define five regions and four transitions (two ridges and two valleys).
Specifically, there is a ridge between $S_0^{}$ and $D_1^{}$,
where we expect the lattice vectors to change continuously, but not smoothly.
A second local maximum within the $S_1^{}$ region is not identified as a transition but according to our theory is a ridge,
which means the two sides must have different intersection equations.

The 1-parameter family of truncated tetrahedra in Ref.~\cite{acs} (see their Fig.~2(a)) 
corresponds to the left edge (and by symmetry also to the bottom edge) of the $323\!\cdot\!4$ family. 
The authors find eight regions and seven transitions, with which we are in agreement: $\{\rho_{60}^{}, \rho_{66}^{}, \rho_{68}^{}, \rho_{31}^{}, \rho_{56}^{},
\rho_{62}^{}, \rho_{41}^{}, \rho_{70}^{}\}$.
We note that their vector length curves are continuous across ridges, 
which corresponds to our ridges $[\rho_{66}^{}\wedge\rho_{68}^{}]$ and $[\rho_{31}^{}\wedge\rho_{56}^{}]$.

\begin{figure*}
\centering
\includegraphics[width=6.75in,height=9in]{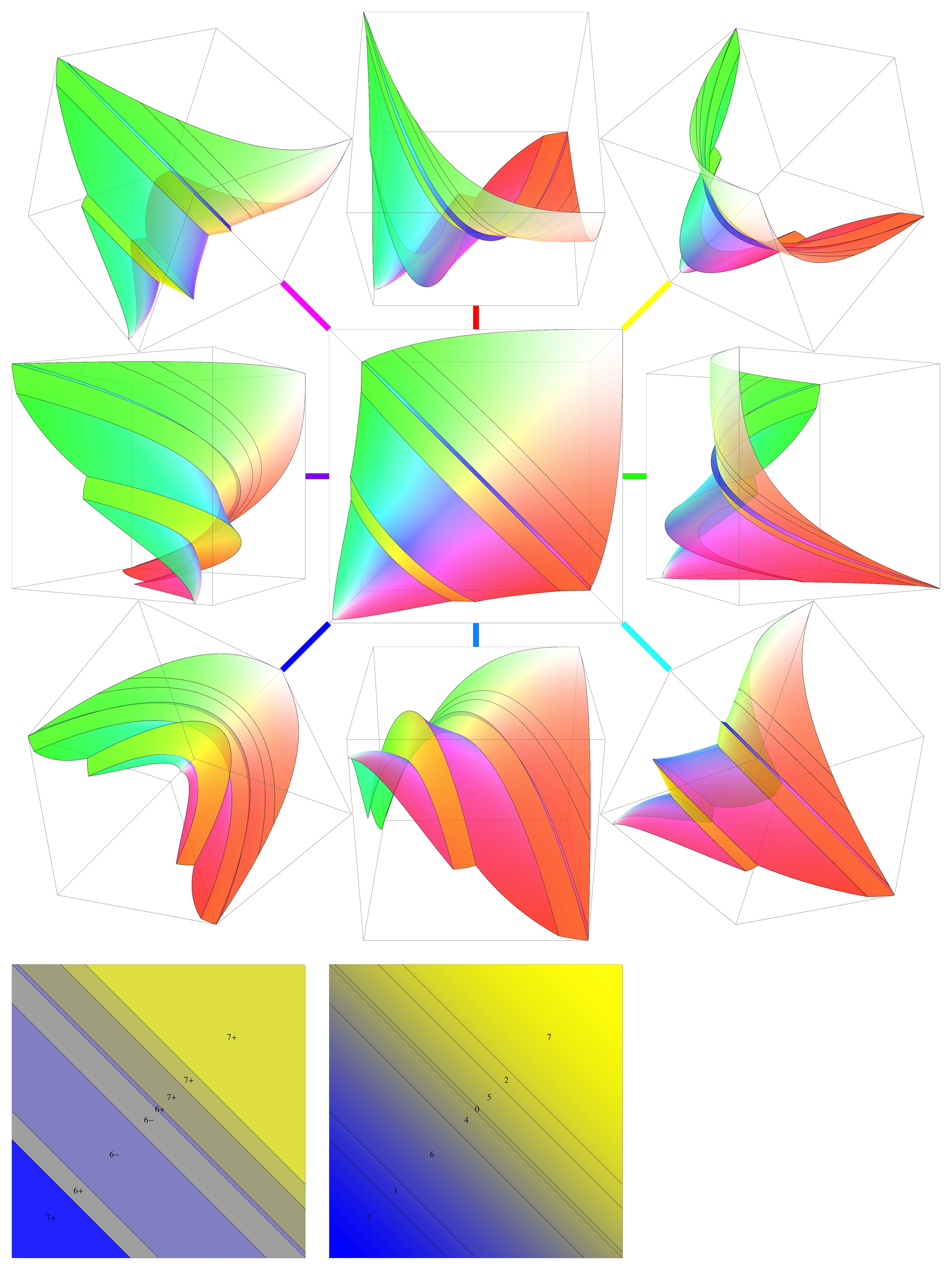}
\caption{$323\!\cdot\!1$ family optimal density surface ($3\times 3$ grid of cubes), parallel contacts (bottom left), face areas and region size (bottom center).}
\label{fig:pack3231}
\end{figure*}

\begin{figure*}
\centering
\includegraphics[width=6.75in,height=9in]{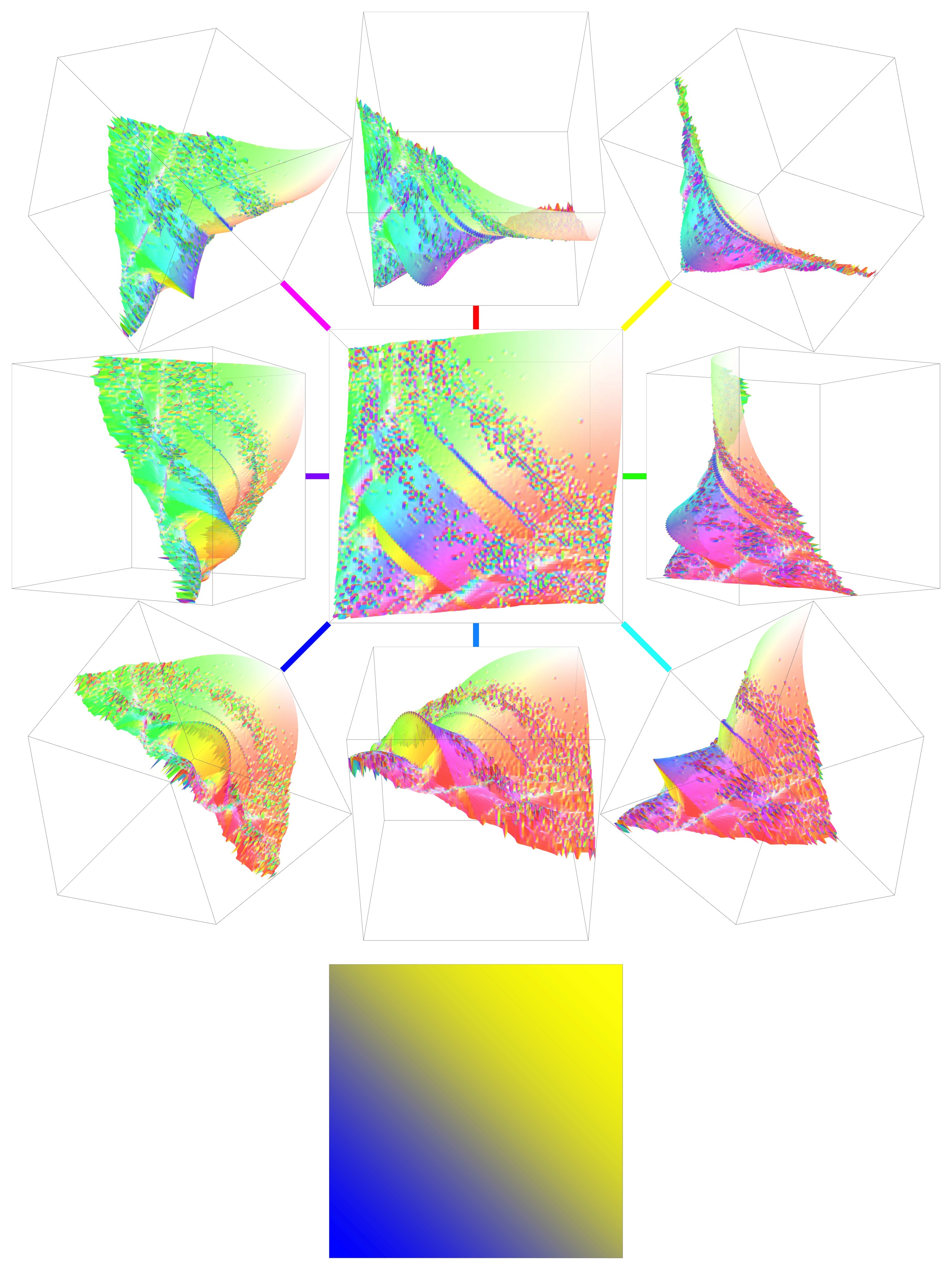}
\caption{$323\!\cdot\!3$ family optimal density surface ($3\times 3$ grid of cubes), face areas (bottom center).}
\label{fig:pack3233}
\end{figure*}

\begin{figure*}
\centering
\includegraphics[width=6.75in,height=9in]{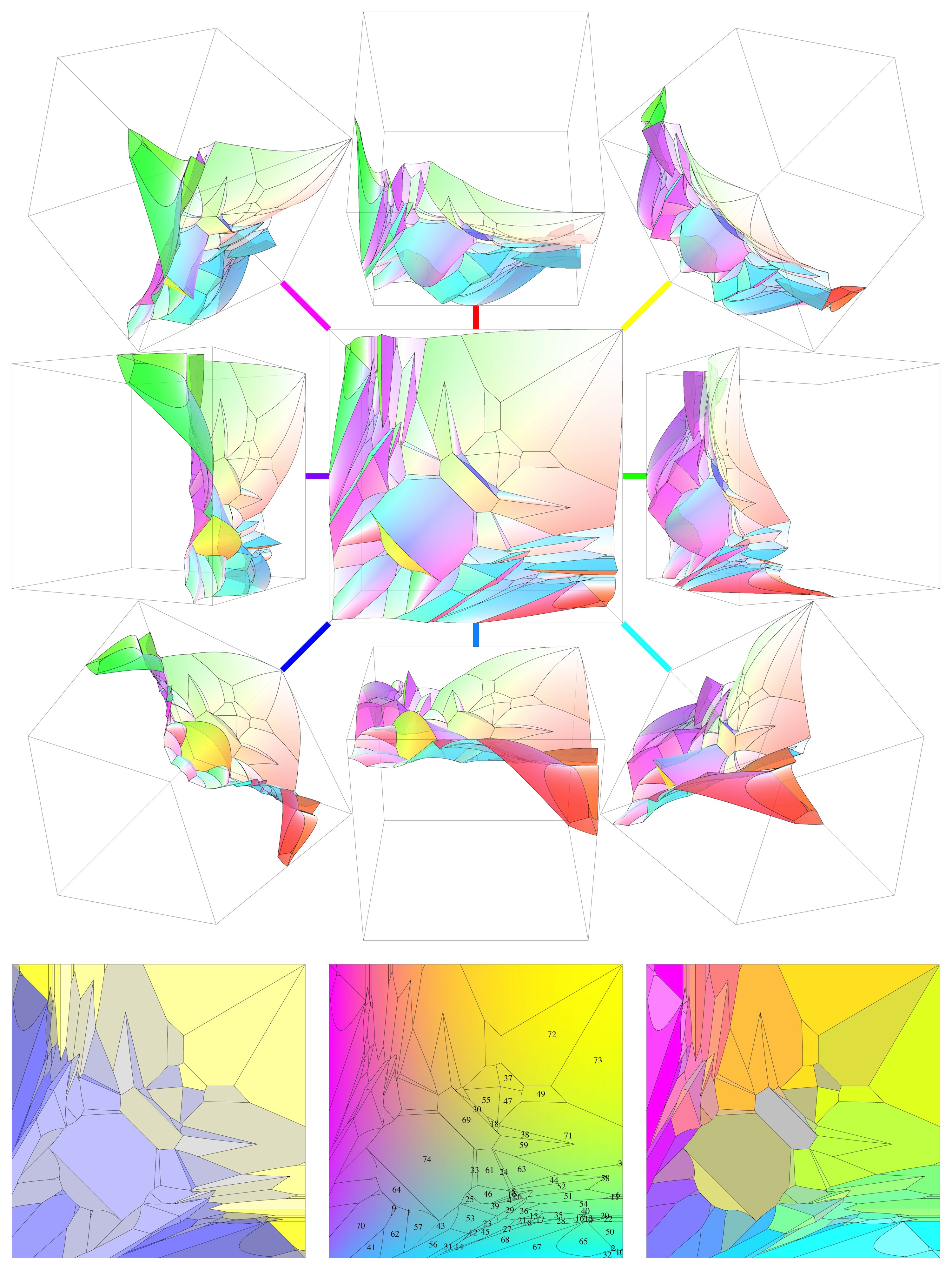}
\caption{$323\!\cdot\!2$ family optimal density surface ($3\times 3$ grid of cubes), parallel contacts (bottom left), anti-parallel contacts (bottom right), face areas and region size (bottom center).}
\label{fig:pack3232}
\end{figure*}

\begin{figure*}
\centering
\includegraphics[width=6.75in,height=9in]{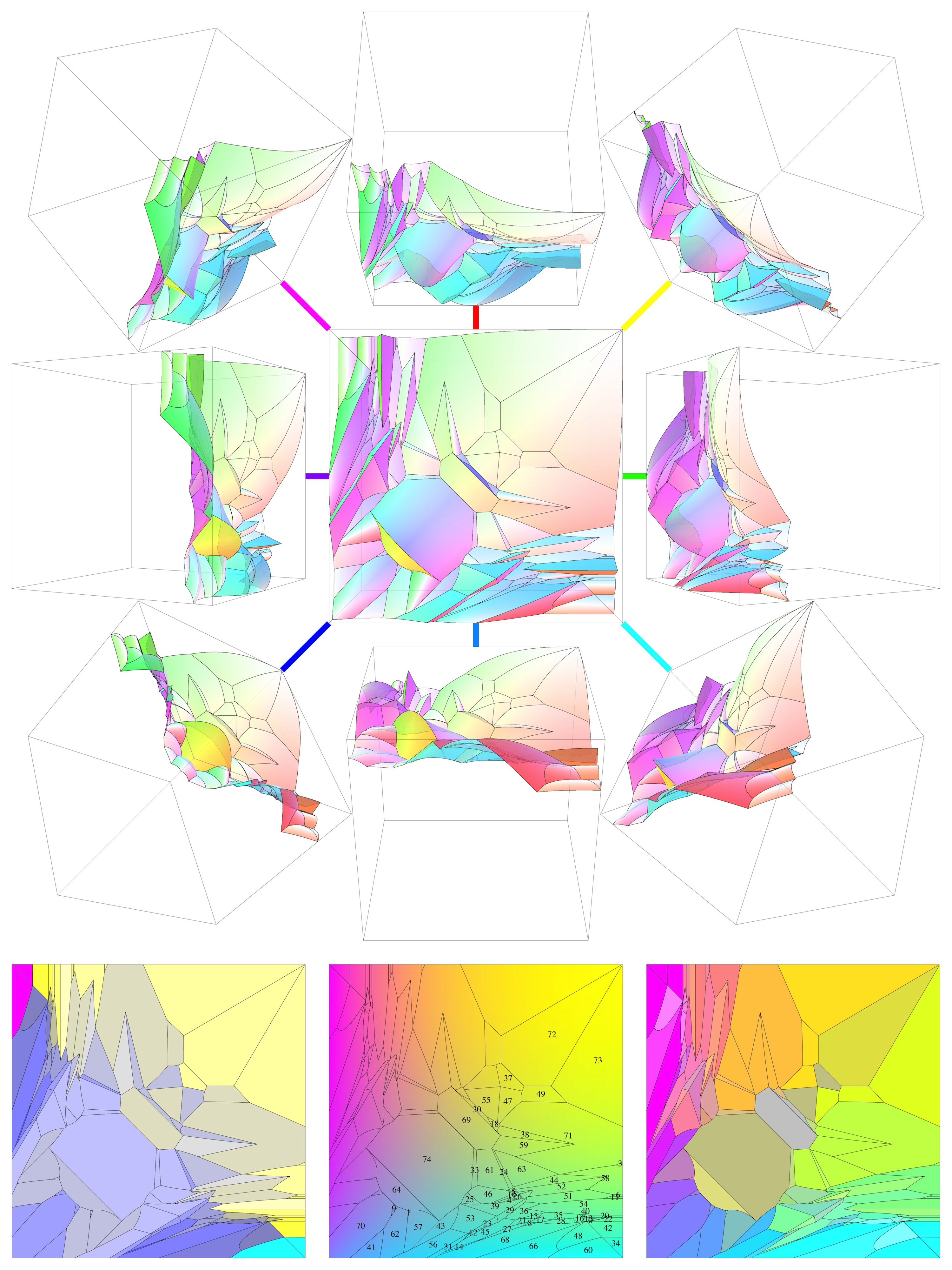}
\caption{$323\!\cdot\!4$ family optimal density surface ($3\times 3$ grid of cubes), parallel contacts (bottom left), anti-parallel contacts (bottom right), face areas and region size (bottom center).}
\label{fig:pack3234}
\end{figure*}

The 1-parameter family of Gantapara \emph{et al.}~\cite{dijkstra13} ranging from the cube to the octahedron via the cuboctahedron corresponds
to the symmetry diagonal in the $323$ family.
While they report 14 regions (see their Fig.~1 and Supplemental Material), we find that some of 
them are identical and instead we have only eight regions (Fig.~\ref{fig:pack3231}):
$\{\rho_{7}^{},\rho_{2}^{},\rho_{5}^{},\rho_0^{},\rho_{4}^{},\rho_{6}^{},\rho_{1}^{},\rho_{3}^{}\}$. 
Our region $\rho_{4}^{}$ corresponds to their regions VI, VII, VIII, 
and our region $\rho_{6}^{}$ corresponds to their IX, X, XI, XII.
The discontinuities in the lattice basis vector lengths and angles that appear due to ordering them by magnitude, were
interpreted by the authors as the boundary X-XI. The discontinuities that appear due to inconsistent choice of basis vectors 
were interpreted as boundaries 
VI-VII, VII-VIII, IX-X, XI-XII.
In addition, our region $\rho_{2}^{}$ corresponds to their regions II, III but we find no boundary in-between. 


\section{Summary and conclusion}

We studied three 2-parameter families of symmetric polyhedra, where the shape is continuously deformed via 
vertex and edge truncations. The surface of maximum packing density was determined as a function of shape parameters $\<a,c\>$. 
We defined an equivalence relation of packings based on the topological type of contacts and intersection equations,
which allowed a classification of packings into regions, as well as a classification of the boundaries between adjacent regions into three types: valley, ridge, and tangent.
Ridges are special boundaries. The lattice deforms continuously across a ridge, but the set of neighbors in contact changes.
We have shown that for centrally symmetric shapes a ridge always separates a region of Minkowski type $G_{}^{6-}$ from a region of 
Minkowski type $G_{}^{6+}$. Right on the ridge, the packing has Minkowski type $G_{}^{7+}$.

Currently, it is not possible to predict based solely on the geometry of the particle whether the surface of 
maximum packing density will be simple or not. 
As in all other investigations of densest packings, we use periodic boundary conditions and
look only at relatively small unit cells; consequently we cannot say unequivocally that no denser packings exist.
It is clear from our results, however, that surfaces of
maximum packing density can exhibit surprising complexity and many `micro-regions', as we have 
demonstrated for the non-centrally symmetric particles of the $323$ family.
Finally, we hope others will build upon our work and extend it to other continuous families of shapes. 

Our results demonstrate that a fuller appreciation of the way in which polyhedra of various shapes can be expected 
to pack at high density can be achieved through exploration of not only the specific shape of immediate interest, 
but also shapes near to it in shape space as achieved by small deformations.  Here we investigated deformations 
achieved via truncations, but other shape anisotropy dimensions may be explored in the same manner~\cite{glotzer07,grega,gregb}. 
This extended knowledge should be especially helpful in the design and synthesis of particles intended to pack into target structures.  
For example, if the particle in question is one whose densest packing lies in a smooth region of the packing surface, then it may be 
more likely to achieve that packing than it would be for a particle whose densest packing lies in a complex region of the surface, 
adjacent to many different nearby packings.  Achieving the latter  would require highly uniform particles, whereas the former may 
tolerate certain shape imperfections.  For these reasons, in the context of particle packings we advocate thinking of a shape 
not as a fixed and static property, but rather as part of a continuum of shape anisotropy  characteristics (dimensions) that are, 
at least in principle, tunable to achieve certain targets.  In this way, some shapes are seen as more ``interesting'' than others because of the 
local topography of their underlying surface of densest packings.

\begin{acknowledgments}
E.R.C.\ acknowledges the National Science Foundation MSPRF grant DMS-1204686.  
D.K.\ acknowledges the FP7 Marie Curie Actions of the European Commission, Grant Agreement PIOF-GA-2011-302490 Actsa.   
The numerical studies of the densest packings were supported by the U.S. Department of
Energy, Office of Basic Energy Sciences, Division of Materials
Sciences and Engineering, Biomolecular Materials Program under Award No. DEFG02-02ER46000.
S.C.G., P.D.\ and M.E.\ were supported by the DOD/ASD(R\&E) under Award No.\ N00244-09-1-0062.
Any opinions, findings, and conclusions or recommendations expressed in this publication are those of the authors and 
do not necessarily reflect the views of the DOD/ASD(R\&E).
\end{acknowledgments}

\clearpage
\bibliography{library.bib}

\end{document}